\documentclass{PoS}

\title{Radiative and EW Penguin $B$ Decays at Belle}

\ShortTitle{Radiative and EW Penguin $B$ Decays at Belle}

\author{\speaker{Akimasa Ishikawa}\\
        Tohoku University\\
        E-mail: \email{akimasa@epx.phys.tohoku.ac.jp}}

\abstract{We report on new results on $B \to K^* \gamma$ and recent studies on $B \to K^* \ell^+ \ell^-$ and $B \to h \nu \bar{\nu}$ at Belle at KEKB accelerator. All the analyses used full data sample of 711~fb${}^{-1}$ taken on $\Upsilon(4S)$ resonance.}

\FullConference{The 15th International Conference on Flavor Physics \& CP Violation\\
		5-9 June 2017\\
		Prague, Czech Republic}

\def\Journal#1#2#3#4{{#1} {\bf #2}, #3 (#4)}

\def\JournalJHEP#1#2#3#4{{#1}, #2 (#3) #4}

\def\PLB{{ Phys. Lett.}  B}
\def\PRL{ Phys. Rev. Lett.}
\def\PRD{{ Phys. Rev.} D}

\def\JHEP{ J. High Energy Phys.}

\def\Mbc{M_{\rm bc}}

\begin{document}

\section{Introduction}
Radiative and electroweak penguin $B$ decays are sensitive to new physics~(NP). These processes are suppressed by Cabbibo-Kobayashi-Maskawa (CKM) matrix elements~\cite{Cabibbo:1963yz,Kobayashi:1973fv}, $V_{ts}$ or $V_{td}$, and a loop factor in the SM. In NP models, unobserved heavy particles might be able to enter in the loop, or might mediate the process even via tree level with comparable amplitudes to the SM ones. Further, these processes are experimentally and theoretically clean due to final states having color singlet leptons or photons. Thus, radiative and electroweak penguin $B$ decays are ideal tools to search for NP.

\section{Evidence for Isospin Violation in $B \to K^* \gamma$}
Radiative $B \to K^* \gamma$ decay proceeds predominantly via one-loop electromagnetic penguin diagrams. This process is also possible via annihilation diagrams; however, the amplitudes are highly suppressed by $\Lambda_{\rm QCD}/m_b$ and CKM matrix elements in the SM~\cite{Bosch:2001gv,Beneke:2000wa}. Since heavy new particles in NP could contributes to penguin diagrams and/or annihilation diagrams, the branching fractions and direct $CP$ violation~($A_{CP}$) might differ from the SM predictions. NP contributions to annihilation diagrams could be different between charged and neutral $B$ mesons, the isospin differences of the decay width~($\Delta_{0+}$) and the $A_{CP}$~($\Delta A_{CP}$) are good probes to NP. The $\Delta_{0+}$, $A_{CP}$ and $\Delta A_{CP}$ are defined as,
\begin{eqnarray}
&&\Delta_{0+}   = \frac{\Gamma(B^0 \to K^{*0} \gamma)-\Gamma(B^+ \to K^{*+} \gamma)}{\Gamma(B^0 \to K^{*0} \gamma)+\Gamma(B^+ \to K^{*+} \gamma)},\\
&&A_{CP}        = \frac{\Gamma(\bar{B} \to \bar{K}^* \gamma)-\Gamma(B \to K^* \gamma)}{\Gamma(\bar{B} \to \bar{K}^* \gamma)+\Gamma(B \to K^* \gamma)},\\
&&\Delta A_{CP} = A_{CP}(B^+ \to K^{*+} \gamma) - A_{CP}(B^0 \to K^{*0} \gamma),
\end{eqnarray}
Predictions of the isospin asymmetry range from 2\% to 8\% with a typical uncertainty of 2\% in the SM~\cite{Keum:2004is,Lyon:2013gba,Ball:2006eu,Kagan:2001zk,Jung:2012vu,Ahmady:2013cva}, while a large deviation from the SM predictions is possible due to NP~\cite{Lyon:2013gba,Kagan:2001zk,Jung:2012vu}. $A_{CP}$ is predicted to be small in the SM~\cite{Keum:2004is,Jung:2012vu,Greub:1994tb,Paul:2016urs}; hence, a measurement of $CP$ violation is a good probe for NP~\cite{Dariescu:2007gr}. The isospin difference of direct $CP$ violation is theoretically discussed in the context of inclusive $B \to X_s \gamma$ process~\cite{Benzke:2010tq} but heretofore not in the exclusive $B \to K^* \gamma$ channel; however, $\Delta A_{CP}$ here will be useful to identify NP once $A_{CP}$ is observed. 

The current world averages of the isospin and direct $CP$ asymmetries are $\Delta_{0+} = (+5.2 \pm 2.6)$\%, $A_{CP}(B^0 \to K^{*0} \gamma) = (-0.2\pm1.5)$\%, $A_{CP}(B^+ \to K^{*+} \gamma) = (+1.8 \pm 2.9)$\% and $A_{CP}(B \to K^{*} \gamma) = (-0.3 \pm 1.7)$\%~\cite{Olive:2016xmw}, respectively, which are consistent with predictions in the SM and give strong constraints on NP~\cite{Jung:2012vu,Paul:2016urs,Altmannshofer:2014rta,DescotesGenon:2011yn,Mahmoudi:2007gd}. The world averages of branching fractions are also consistent with predictions within the SM~\cite{Bosch:2001gv,Keum:2004is,Ball:2006eu,Jung:2012vu,Greub:1994tb,Straub:2015ica,Ali:2007sj,Ali:2001ez} and are used for constraining NP~\cite{Jung:2012vu,Paul:2016urs,Ciuchini:2016weo}. This analysis supersedes our previous publication~\cite{Nakao:2004th}.

We reconstruct $B^0 \to K^{*0} \gamma$ and $B^+ \to K^{*+} \gamma$ decays, where $K^*$ is formed from $K^+ \pi^-$, $K_S^0 \pi^0$, $K^+ \pi^0$ or $K_S^0 \pi^+$ combinations. The dominant background from continuum events is suppressed using a multivariate analysis with a neural network~\cite{NB} using inputs of event shape, kinematic, and flavor tagging quality variables.

\begin{figure}[htb]
\begin{center}
\includegraphics[width=0.35\textwidth]{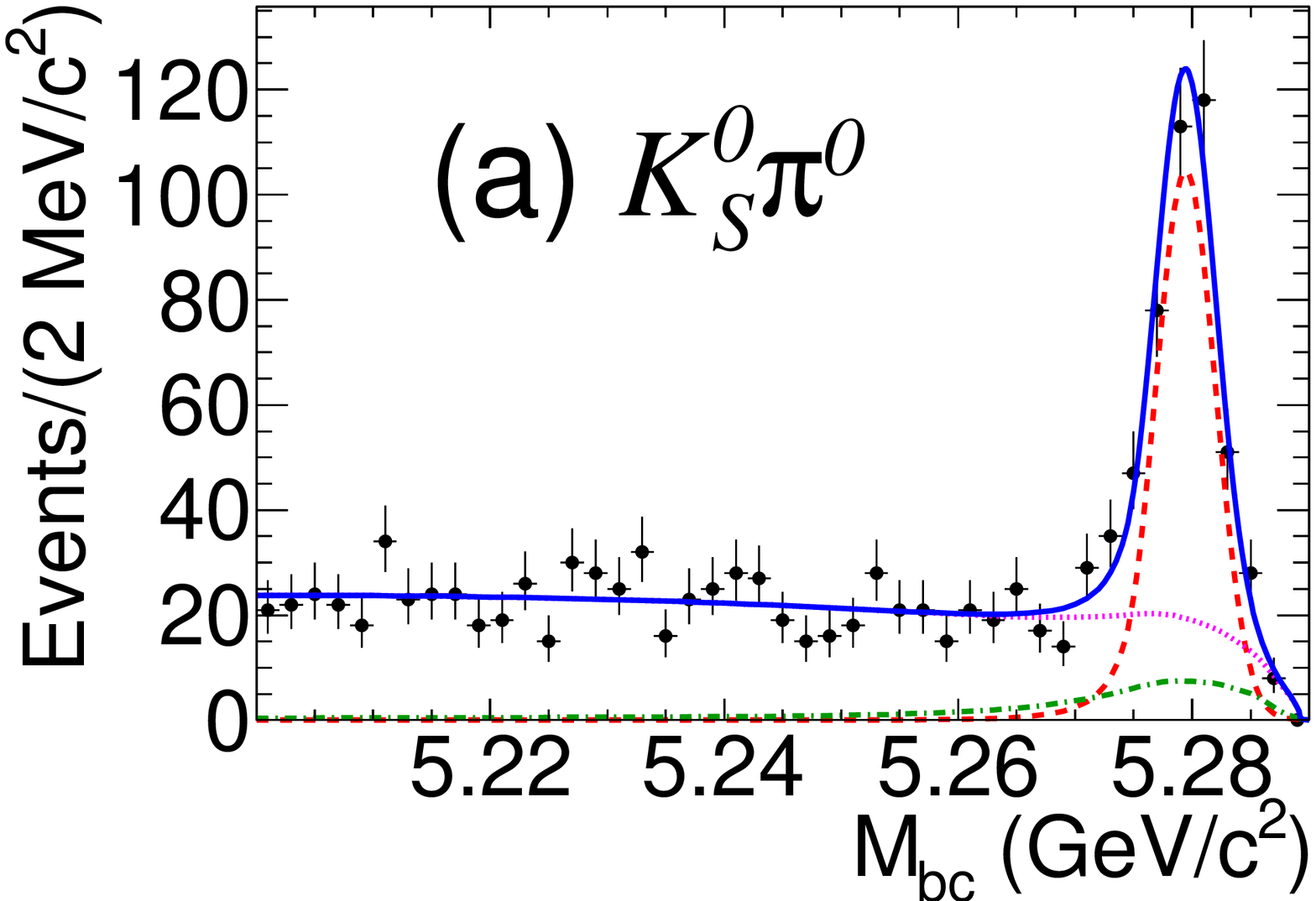}\\
\includegraphics[width=0.35\textwidth]{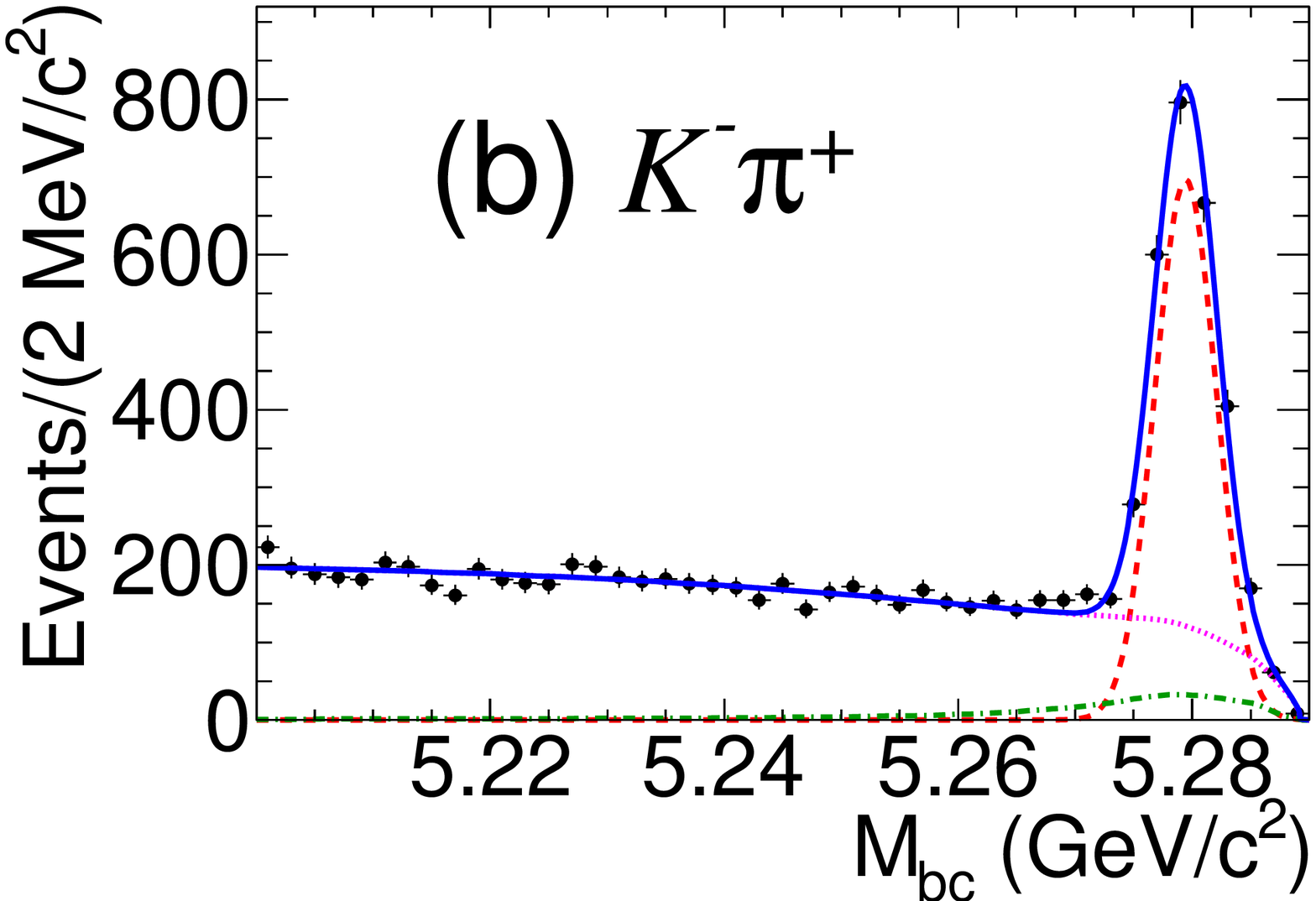}
\includegraphics[width=0.35\textwidth]{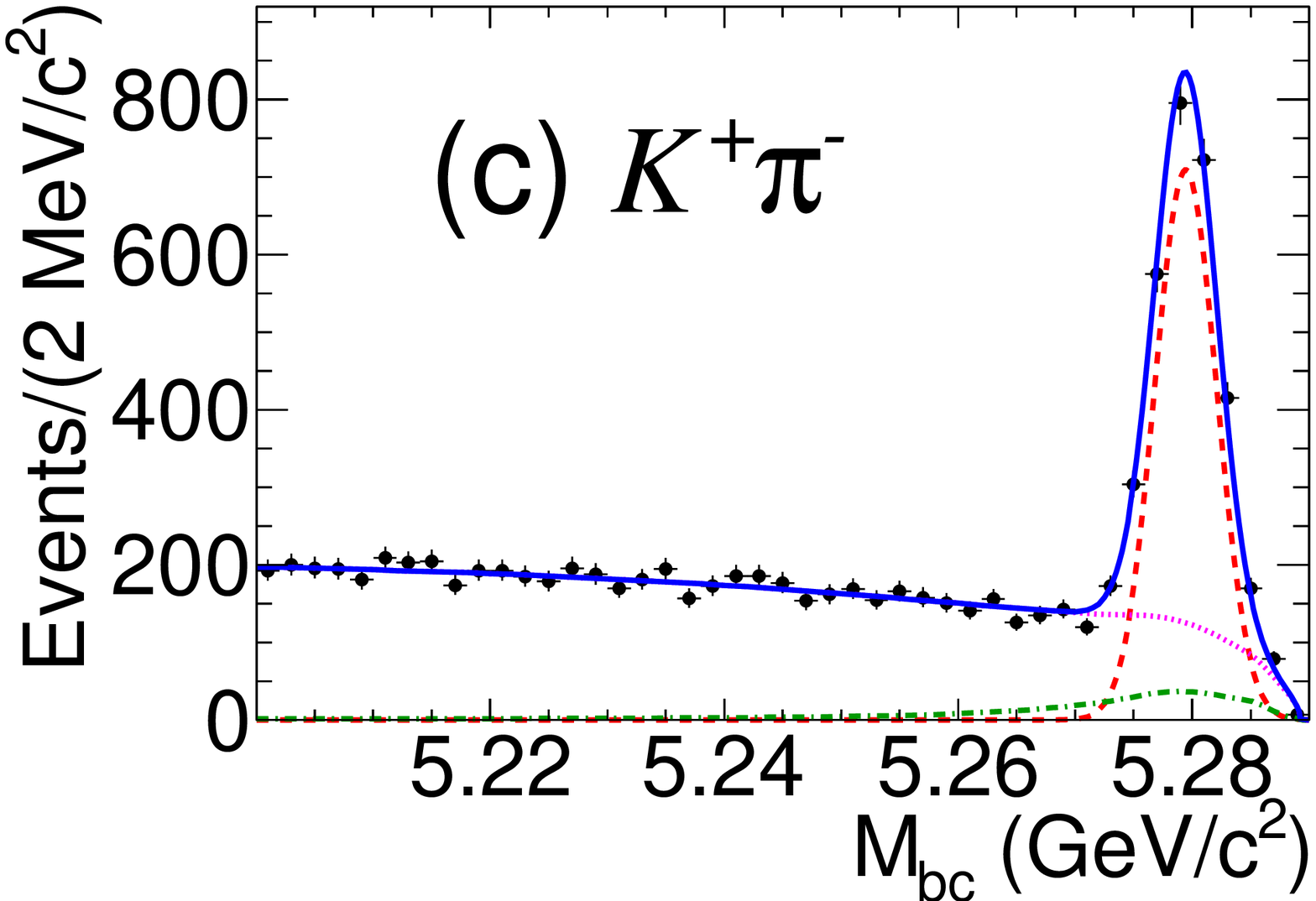}\\
\includegraphics[width=0.35\textwidth]{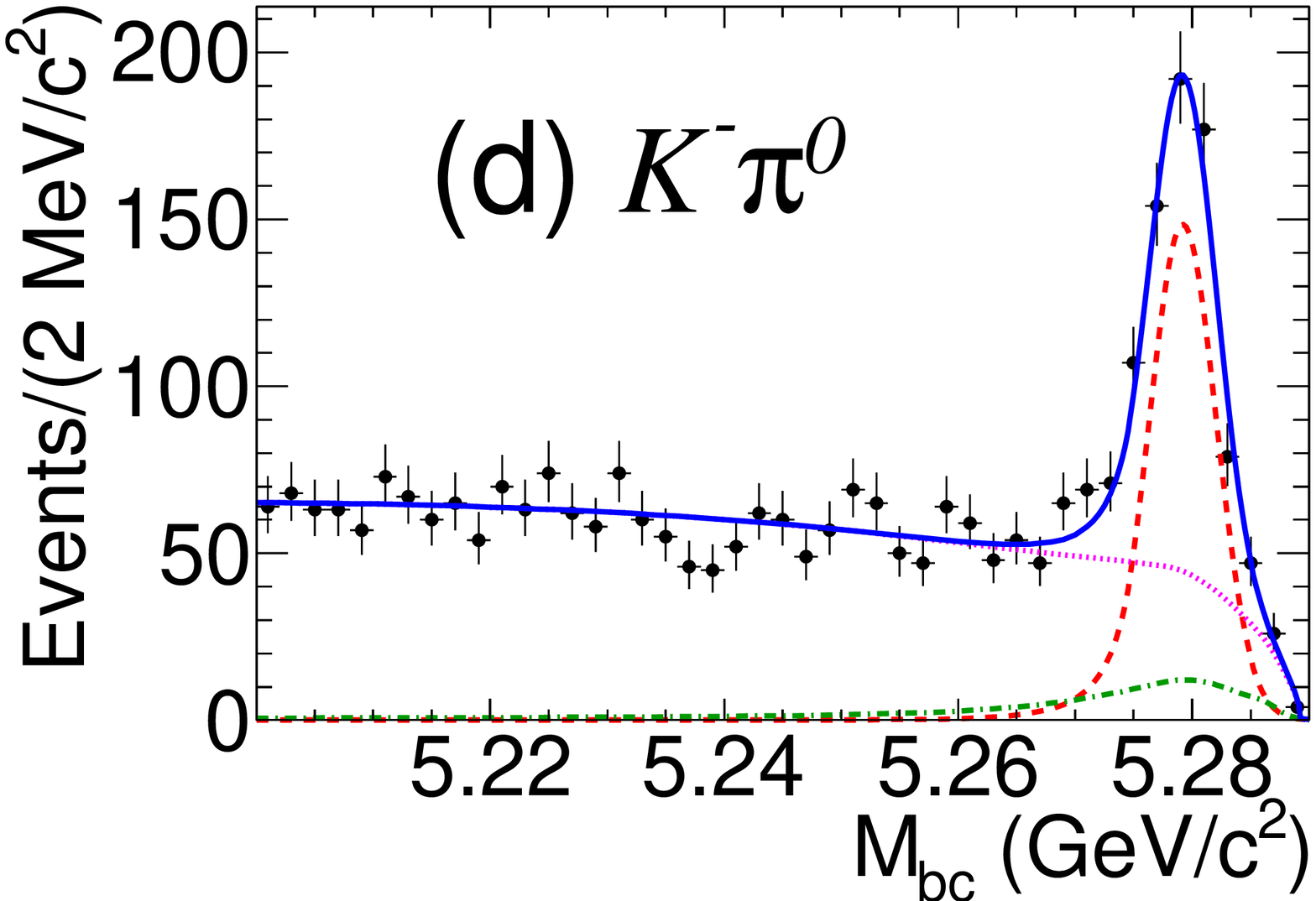}
\includegraphics[width=0.35\textwidth]{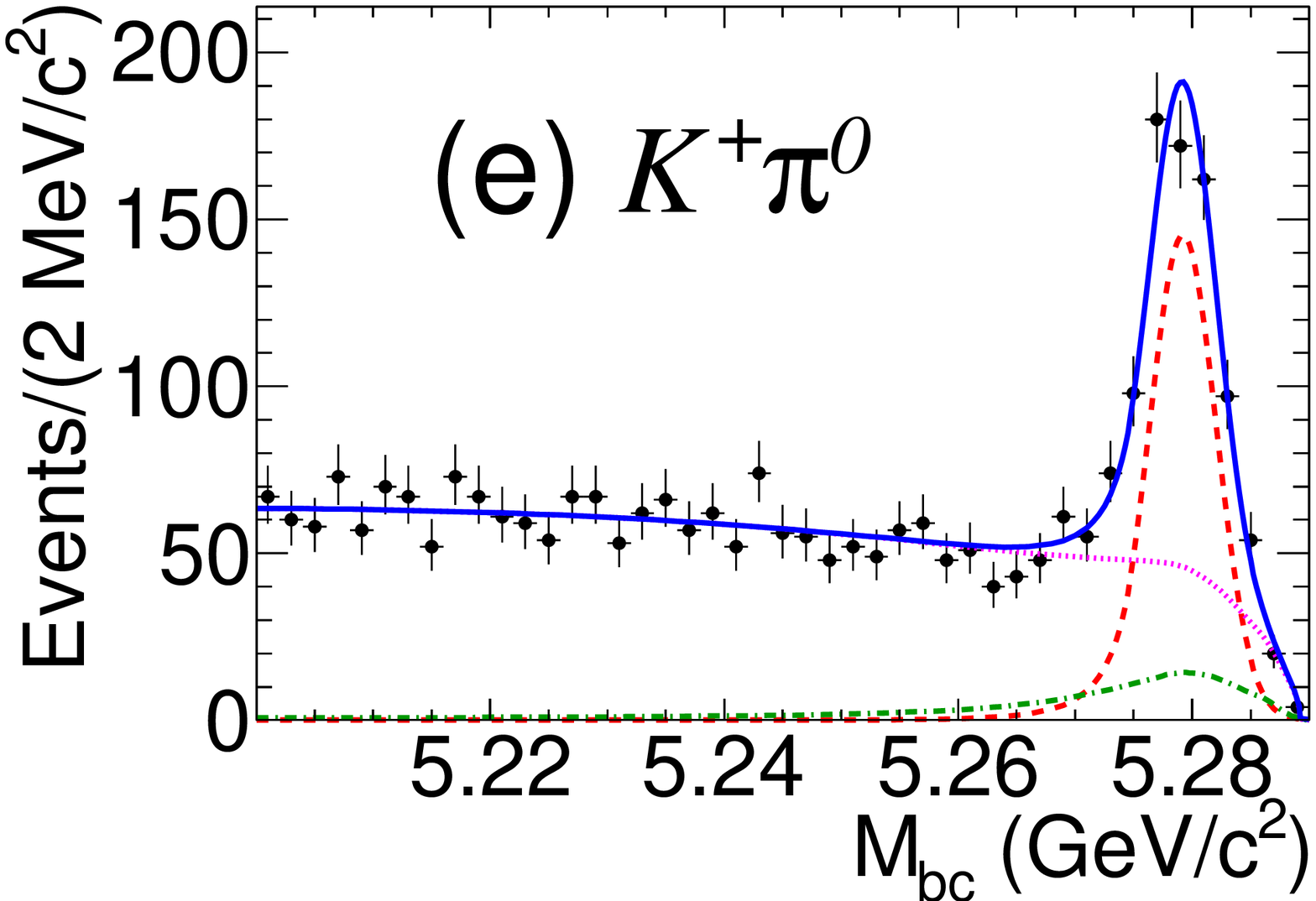}\\
\includegraphics[width=0.35\textwidth]{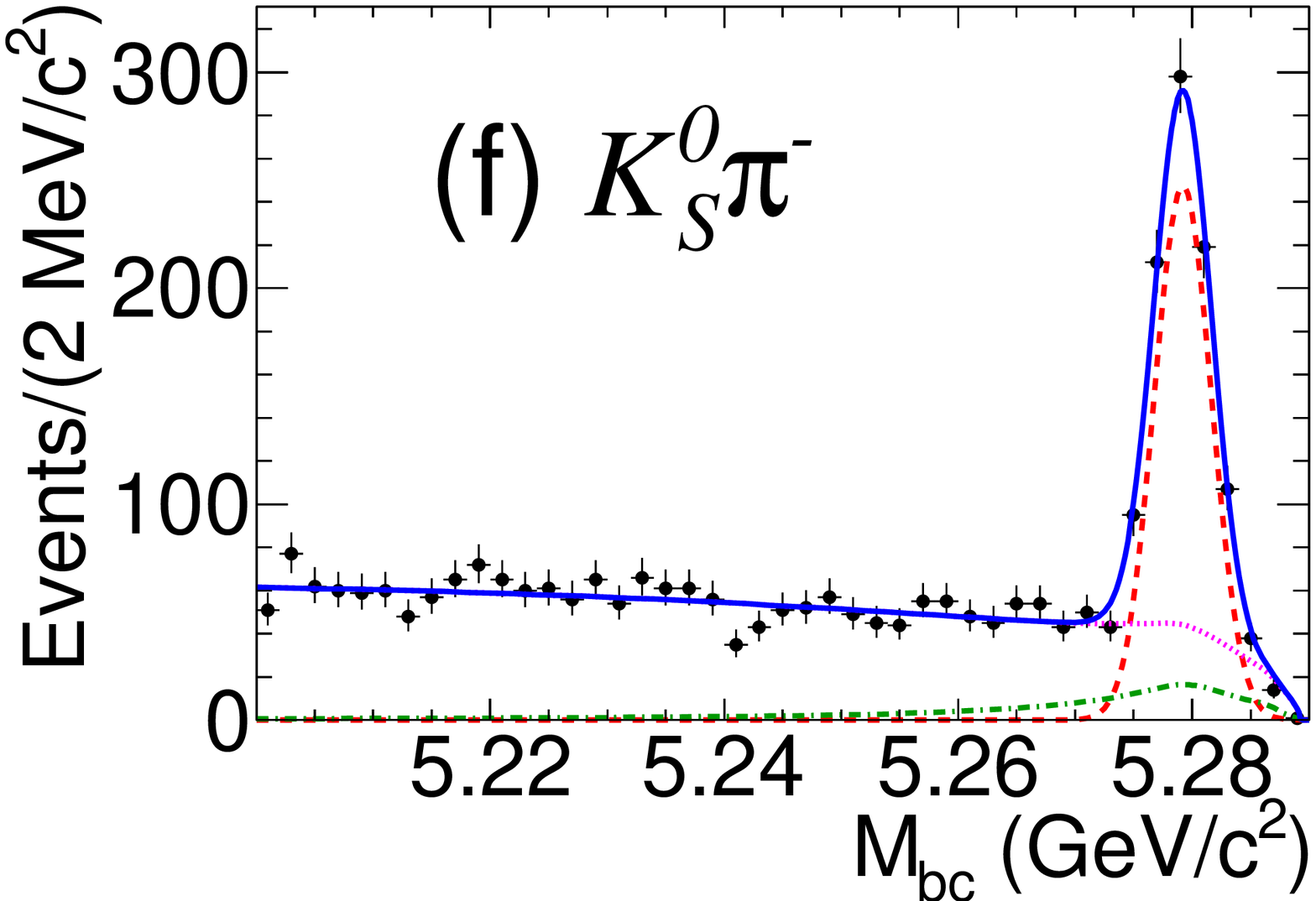}
\includegraphics[width=0.35\textwidth]{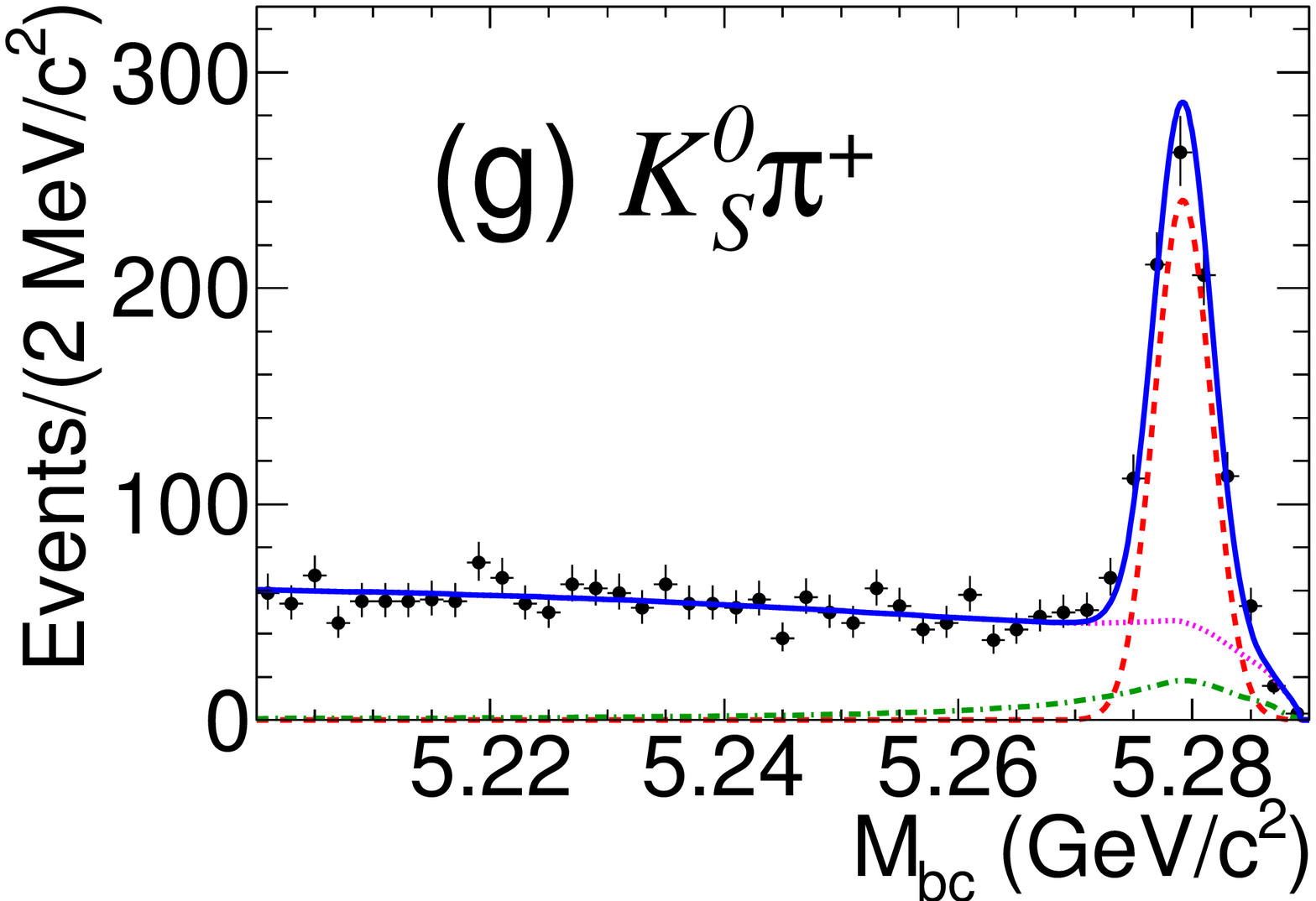}\\
\caption{$\Mbc$ distributions for (a)~$K_S^0\pi^0$, (b)~$K^-\pi^+$, (c)~$K^+\pi^-$, (d)~$K^-\pi^0$ (e)~$K^+\pi^0$, (f)~$K_S^0\pi^-$ and (g)~$K_S^0\pi^+$. The points with error bars show the data, the dashed (red) curves represent signal, the dotted-dashed (green) curves are $B\bar{B}$ background, the dotted (magenta) curves show total background, and solid (blue) curves are the total.}
\label{fig:mbc}
\end{center}
\end{figure}

To determine the branching fractions and direct $CP$ asymmetries as well as $\Delta A_{CP}$ and $\Delta_{0+}$, we perform extended unbinned maximum likelihood fits to the seven $\Mbc$ distributions~(Fig.~\ref{fig:mbc}) and the results are
  \begin{eqnarray*}
{\cal{B}}(B^0 \to K^{*0} \gamma) &=& (3.96 \pm 0.07 \pm 0.14)\times10^{-5},\\ 
{\cal{B}}(B^+ \to K^{*+} \gamma) &=& (3.76 \pm 0.10 \pm 0.12)\times10^{-5},\\ 
A_{CP}(B^0 \to K^{*0} \gamma) &=& (-1.3 \pm 1.7 \pm 0.4)\%,\\ 
A_{CP}(B^+ \to K^{*+} \gamma) &=& (+1.1 \pm 2.3 \pm 0.3)\%,\\ 
A_{CP}(B \to K^{*} \gamma)    &=& (-0.4 \pm 1.4 \pm 0.3)\%,\\ 
\Delta_{0+}   &=& (+6.2 \pm 1.5 \pm 0.6 \pm 1.2)\%, \\
\Delta A_{CP} &=& (+2.4 \pm 2.8 \pm 0.5)\%, 
  \end{eqnarray*}
where the first uncertainty is statistical, the second is systematic, and the third for $\Delta_{0+}$ is due to the uncertainty in $f_{+-}/f_{00}$~\cite{Horiguchi2017}. 
We find evidence for isospin violation in $B \to K^* \gamma$ decays with a significance of 3.1$\sigma$, and this result is consistent with the predictions in the SM~\cite{Keum:2004is,Lyon:2013gba,Ball:2006eu,Kagan:2001zk,Jung:2012vu,Ahmady:2013cva,Greub:1994tb}. The $A_{CP}$ and $\Delta A_{CP}$ values are consistent with zero. All the measurements are the most precise to date and will be used for constraining parameter space in NP models.

We also calculate the ratio of branching fractions of $B^0 \to K^{*0} \gamma$ to $B_s^0 \to \phi \gamma$, which is sensitive to annihilation diagrams~\cite{Lyon:2013gba}, based on the branching fraction measurement reported here and the Belle result for the ${\cal{B}}(B_s^0 \to \phi \gamma)$~\cite{Dutta:2014sxo}. To cancel some systematic uncertainties, we take only the $K^+\pi^-$ mode for the branching fractions for $B^0 \to K^{*0} \gamma$. The result is
\begin{eqnarray*}
\frac{{\cal{B}}(B^0 \to K^{*0} \gamma)}{{\cal{B}}(B_s^0 \to \phi \gamma)} &=& 1.10\pm0.16\pm0.09\pm0.18,
\end{eqnarray*}
where the first uncertainty is statistical, the second is systematic, and the third is due to the fraction of $B_s^{(*)0}\bar{B}_s^{(*)0}$ production in $\Upsilon(5S)$ decays.
This result is consistent with predictions in the SM~\cite{Lyon:2013gba,Ali:2007sj}.

\section{Lepton Flavor Dependent Angular Analysis of $B \to K^* \ell^+ \ell^-$}
The $b \to s \ell^+ \ell^-$ decays were observed by Belle Collaboration about 15 years before~\cite{bib:b2sllBelle} which opened new door to search for NP. The BF and forward-backward asymmetry as functions of $q^2$ in $B \to K^* \ell^+ \ell^-$ are important observables for NP searches, and several experiments already measured~\cite{bib:AFBex}. Full angular analysis of $B \to K^* \ell^+ \ell^-$ with optimized observables~\cite{bib:S5pred}, which are less sensitive to form factor uncertainties, are very powerful tools to search for NP. LHCb first reported the results~\cite{bib:LHCbP52013} and one of the observable, $P_5'$, is deviated about 3.4~$\sigma$ from a prediction in the SM by DHMV~\cite{bib:DHMV} (There is a discussion in theory community that the deviation might be able to be explained by charm-loop~\cite{bib:JM,bib:BSZ,bib:charmloop}). This could indicate NP contributions in the $b \to s \ell^+ \ell^-$ process. Lepton flavor universality holds in the SM. The ratios of the branching fractions of $B \to K^{(*)} \mu^+ \mu^-$ to $B \to K^{(*)} e^+ e^-$~($R_{K^{(*)}}$) as a function of $q^2$ are unity except for very low $q^2$ region due to finite lepton mass effect. The $R_{K^{(*)}}$ measured by LHCb are deviated from the SM predictions about 2.6$\sigma$~\cite{Aaij:2014ora,Aaij:2017vbb} and these could also suggest NP in the process. 
By a global fit to observables in $b \to s \gamma$ and $b \to s \ell^+ \ell^-$ including $P_5'$ and $R_{K^{(*)}}$, one of the Wilson coefficients for muon, $C_{9\mu}$, is deviated about -1 from the SM prediction (or $C_{9\mu}^{\rm NP}=-C_{10\mu}^{\rm NP}\sim-0.6$) while the same for electron is consistent with the SM~\cite{bib:global}. Thus, next analysis which should be performed is lepton flavor universality in angular observables.

\begin{figure}
\begin{center}
\includegraphics[width=.45\textwidth]{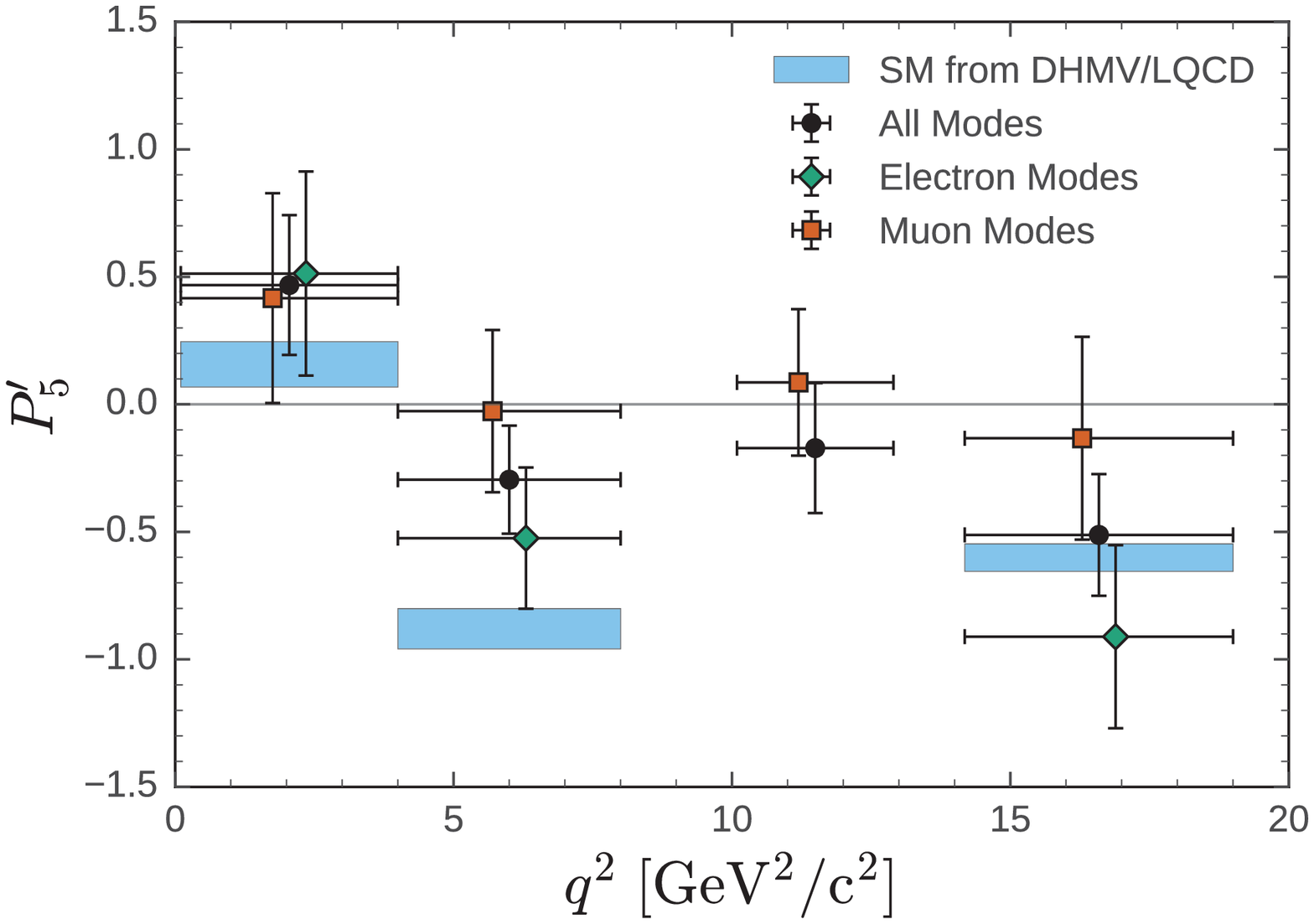}
\includegraphics[width=.45\textwidth]{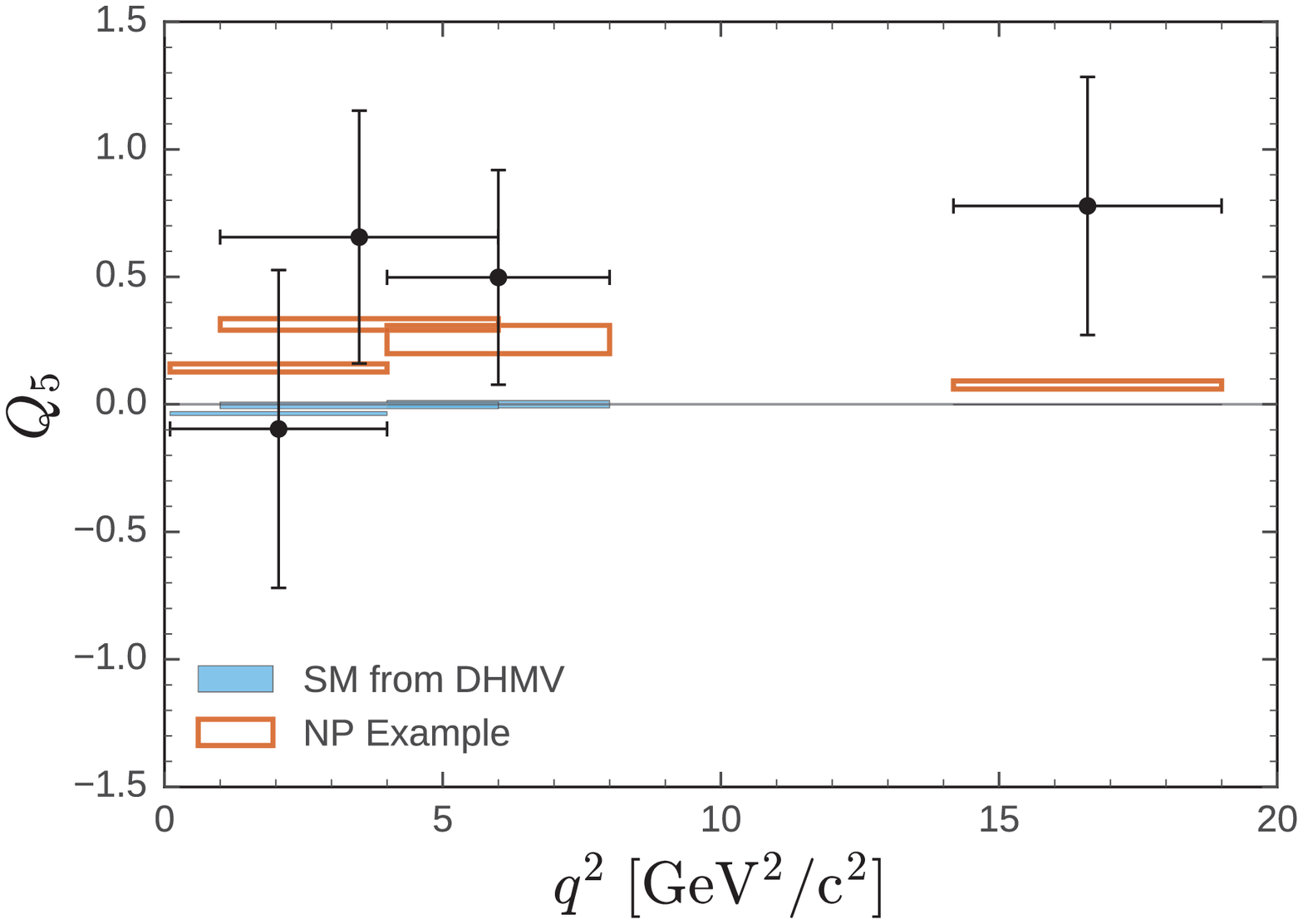}
\end{center}
\caption{$P_5'$~(left) and $Q_5$~(right) distributions in $B \to K^{\ast} \ell^+ \ell^-$.}
\label{fig:p5q5}
\end{figure}

We measured the optimized observables $P_{4,5}'$ using charged and neutral $B \to K^* \ell^+ \ell^-$ decays separately for electron and muon modes, and then took the difference, $Q_i = P_i^{\mu} - P_i^{e}$~\cite{Capdevila:2016ivx}. Even with full data, we expected only 300 signal events which is about 10 times smaller than that at LHCb, the selection criteria should be optimized better than previous analysis~\cite{Wei:2009zv}. We adopted neural net based analysis to select signal candidates and to suppress backgrounds. Signal is extracted by fitting to $M_{\rm bc}$ distributions. We observed $127\pm15$ and $185\pm17$ signal events for electron and muon modes, respectively. For full angular analysis, we adopted the folding method on angular variables, $\theta_{\ell}$, $\theta_K$ and $\phi$, to extract optimized observables which LHCb performed in 2013~\cite{bib:LHCbP52013}. The fit results for $P_5'$ for electron, muon and combined cases are shown in Fig.\ref{fig:p5q5} (left)~\cite{bib:P5}. For combined case, $P_5'$ for $4 < q^2 < 8$ is about 2.5~$\sigma$ deviated from a prediction by DHMV~\cite{bib:DHMV} and is consistent with LHCb result~\cite{bib:LHCbP5}. The results for $Q_5$~(Fig.\ref{fig:p5q5} (right)) are consistent with both the SM and the case for $C_{9\mu}^{\rm NP} = -1.1$. Other observables, $P_4'$ and $Q_4$, are consistent with the SM predictions within errors.

\section{Search for $B \to h \nu \bar{\nu}$}
The di-neutrino emission processes, $B \to h \nu \bar{\nu}$, are not observed yet~\cite{bib:b2snunuBabar2013,bib:b2snunuBelle2013}. This loop process is theoretically interesting since clean prediction is possible thanks to exact factorization and no contributions from charm-loop diagrams~\cite{bib:b2snunupred,bib:b2dnunupred}, and NP effects, such as $C_9$ deviation, could be correlated with $b \to s \ell^+ \ell^-$ in some NP models~\cite{Buchalla:2000sk}. Combined analysis of $b \to s \nu \bar{\nu}$ and $b \to s \ell^+ \ell^-$ allows for new physics test with less form factor uncertainties~\cite{Bartsch:2009qp}. Studies of these process can be also used for searches for new light invisible particles~($X_{\rm inv}$) $B \to h X_{\rm inv}$ or $B \to h X_{\rm inv} \bar{X}_{\rm inv}$~\cite{bib:inv}.
Previous search at Belle used hadronic $B$ tagging~\cite{bib:b2snunuBelle2013} while new measurement used semileptonic tagging.

\begin{figure}
\begin{center}
\includegraphics[width=.45\textwidth]{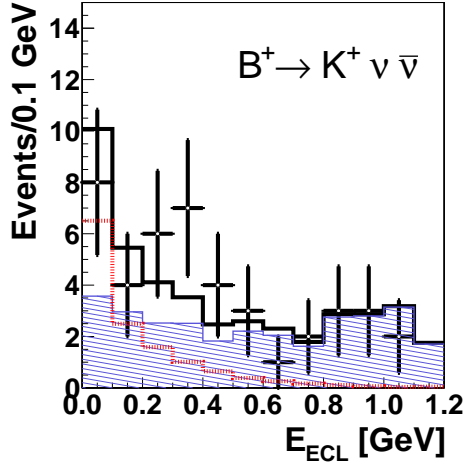}
\end{center}
\caption{$E_{\rm ECL}$ distribution for $B^+ \to K^+ \nu \bar{\nu}$.}
\label{fig:knunu}
\end{figure}

We searched for the $B \to h \nu \bar{\nu}$ decays, where hadronic systems are $\pi^0$, $\pi^+$, $K^0_S$, $K^+$, $\rho^0$, $\rho^+$, $K^{*0}$ or $K^{*+}$. We reconstructed 108 exclusive semileptonic $B$ decays as tagging side. Then, we required momentum of $h$ candidates in the center of mass frame to be $2.96$~GeV~$ > p_h > 0.50$~GeV. To remove misidentified leptons from pions, invariant mass of $K$~($K^*$) and tag-side lepton is required to be far from the $D$ mass region. To suppress the continuum backgrounds, neutral net with input of event shape and kinematic variables was used. The selection was optimized to maximize the figure-of-merit.
We chose extra energy in electromagnetic caloriemeter~($E_{\rm ECL}$) as final discriminator as shown in Fig.~\ref{fig:knunu}. Since the distributions are consistent with background, we set upper limits on the decays as summarized in Tab.~\ref{tab:result}. We obtained world best limits for $K_S^{0}$, $K^{*0}$, $\pi^{+}$, $\pi^{0}$, $\rho^{0}$, $\rho^{+}$~\cite{bib:b2snunuBelle2017}. The limits on BFs for combined $K^*$ modes are just 2.8 times larger than theoretical predictions in the SM~\cite{bib:b2snunupred}, thus Belle II can observe the decay modes, and can measure the BF and longitudinal polarization of $K^*$ with expected precisions of 10\% and 20\%, respectively~\cite{bib:b2tipb2snunu}.

\begin{table}
\begin{center}
\caption{95\% C.L. upper limits and predictions on the BFs for $B \to h \nu \bar{\nu}$ in units of $10^{-6}$. The predictions for $K_S^0$ and $K^{\ast+}$ are obtained from the ones for $K^+$ and $K^{\ast0}$ corrected by lifetime ratio of charged to neutral $B$ mesons.}
\label{tab:result}
\begin{tabular}{l|ccc}
\hline
$h$       & Had. tag~\cite{bib:b2snunuBelle2013} & SL tag~\cite{bib:b2snunuBelle2017} & predictions~\cite{bib:b2snunupred} \\
\hline
$K^+$     & 55 & 19 & $ 3.98 \pm 0.43 \pm 0.19$\\
$K_S^0$   & 97 & 13 & $ 1.85 \pm 0.20 \pm 0.09$\\
$K^{\ast+}$ & 40 & 61 & $ 9.89 \pm 0.93 \pm 0.54 $\\
$K^{\ast0}$ & 55 & 18 & $ 9.19 \pm 0.86 \pm 0.50 $\\
$\pi^+$   & 98 & 14 & --\\
$\pi^0$   & 69 &  9 & --\\
$\rho^+$  & 213 & 30 & --\\
$\rho^0$  & 208 & 40 & --\\
$\phi$    & 127 & -- & --\\
\hline
$K$       & -- & 16 & $ 3.84 \pm 0.41 \pm 0.18 $ \\
$K^{\ast}$  & -- & 27 & $ 9.54 \pm 0.89 \pm 0.52 $ \\
$\pi$     & -- &  8 & -- \\
$\rho$    & -- & 28 & -- \\
\hline
\end{tabular}
\end{center}
\end{table}

\section{Summary}
We have studies radiative and electroweak penguin processes with full data set at Belle experiment. 
We observed evidence for isospin violation in $B \to K^* \gamma$ decay for the first time.
The measured $P_5'$ observable for $4 < q^2 < 8$ is deviated about 2.5~$\sigma$ from the SM prediction by DHMV while the $Q_5$ is consistent with both the SM and NP case for $C_{9\mu}^{\rm NP} = -1.1$. 
The obtained upper limit for $B \to K^* \nu \nu$ is just 2.7 times larger than theoretical prediction, thus Belle II observe the decay modes and can measure the BF and the longitudinal polarization of $K^*$.

\section*{Acknowledgments}
A.~Ishikawa is supported by JSPS Grant Number 16H03968.

\end{document}